\documentstyle[prl,aps,epsf,multicol]{revtex}
\newcommand{\good}[1]{\mbox{\boldmath $#1$}}
\newcommand{\goodsub}[1]{\mbox{\boldmath $\scriptstyle #1$}}
\newcommand{\hlf}{\textstyle \frac{1}{2}}

\begin{document}
\draft

\title{The Electron Spectral Function in Two-Dimensional
Fractionalized Phases} 
\author{C. Lannert$^1$, Matthew P. A. Fisher$^2$, and T. Senthil$^2$}
\address{$^1$Department of Physics, University of California, Santa
Barbara, CA 93106 \\ 
$^2$Institute for Theoretical Physics, University of California, 
Santa Barbara, CA 93106--4030}  

\date{\today}
\maketitle

\begin{abstract}
We study the electron spectral function of various zero-temperature
spin-charge 
separated phases in two dimensions. In these phases, the electron is not a
fundamental excitation of the system, but rather ``decays'' into a
spin-1/2  chargeless fermion (the spinon) and a spinless charge $e$
boson (the chargon). 
Using low-energy effective theories for the spinons (d-wave
pairing plus possible N\'{e}el order), and the chargons (condensed
or quantum disordered bosons), we explore three phases of possible
relevance to the cuprate superconductors: 1) $AF^*$, a fractionalized
antiferromagnet where the spinons are paired into a state with
long-ranged N\'{e}el order and the 
chargons are 1/2-filled and (Mott) insulating, 2) the nodal liquid, a
fractionalized insulator where the spinons are d-wave paired and the
chargons are uncondensed, and 3) the d-wave superconductor, where
the chargons are condensed and the spinons retain a d-wave gap.
Working within the $Z_2$ gauge theory of such fractionalized phases, our
results should be valid at scales below the vison gap. However, on a
phenomenological level, our results should apply to any spin-charge
separated system where the excitations have these low-energy effective
forms. Comparison with ARPES data in the undoped, pseudogapped, and
superconducting regions is made.
\end{abstract}
\vspace{0.15cm}

\begin{multicols}{2}
\narrowtext 

\section{introduction}

Ideas of spin-charge separation have long been considered 
in relation to the cuprate high-$T_c$ materials following P.W. Anderson's
original suggestions \cite{A87}. Phenomenologically, the assumption
that the electron ``breaks apart'' leads to fairly simple explanations
for some otherwise puzzling aspects of these materials. Attempts to
formulate 
this rather elegant idea into a well-defined theory of electrons
living in two or more spatial dimensions have historically been
plagued with problems.  
A recently introduced $Z_2$ gauge theory of strongly correlated
electron systems \cite{Z2} indeed contains both spin-charge
separated and spin-charge confined phases, and we work here within
this formulation. 

\begin{figure}
\epsfxsize=3.5in
\centerline{\epsffile{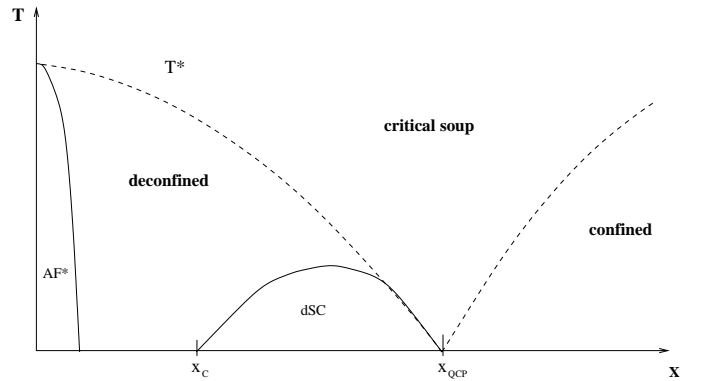}}
\vspace{0.15in}
\caption{Schematic phase diagram for the high $T_c$ cuprates.}
\vspace{0.15in}
\label{phases}
\end{figure}

Among the host of puzzling experimental properties of these materials,
we wish to concentrate here on angle resolved photoemission
spectroscopy (ARPES) experiments which in recent years have reached an
unprecedented level of resolution. With this increased clarity of data
has come increased confusion in theoretical interpretation. In
particular, it seems quite difficult to explain the ARPES
lineshape in the pseudogap regime within Fermi liquid theory. In fact,
any conventional quasiparticle description would seem to predict a
sharp peak in the spectral function,
$A(\good{k} ,\omega)$,
at $\omega(\good{k})$ for some \good{k} in the Brillouin zone. The
data in the 
underdoped compounds in their non-superconducting state, on the other
hand, show only broad and sometimes step-like features. Increased
energy and momentum resolution has made the contrast with the
superconducting state, where a sharp peak does emerge, more striking
and has led to further doubts about the quasiparticle description of
the pseudogap state. As argued elsewhere \cite{QC}, this contrast
between the pseudogap and superconducting lineshapes suggests
that the pseudogap region could be dominated by a
zero-temperature fractionalized phase. In addition, recent results in
the superconducting state  
suggest a connection between the weight under the superconducting
quasiparticle peak and the condensate density \cite{ZXC}. This result
seems rather mysterious from a Fermi-liquid  
point of view, but, as we later show, may have a simple explanation in
terms of separated spin and charge degrees of freedom.
ARPES experiments on undoped compounds also show broad spectral
features rather than well-defined quasiparticle peaks, which has led
us to consider the possibility of a fractionalized antiferromagnet,
dubbed $AF^*$.   
However, the spectral function does show signs of
``sharpening up'' as the system is overdoped, suggesting that there
may be a  
quantum confinement critical point in the cuprate phase diagram, as
shown in Figure \ref{phases}. 

We wish here to explore in more detail the consequences of these
spin-charge separation ideas
for the single-electron spectral function of the cuprate materials at
low doping. 
Working with a fairly simple theory of low-energy spin and charge
excitations in a 
fractionalized phase, we will find qualitative agreement with ARPES
data in the pseudogap and superconducting phases, as well as in the
undoped insulator. Although the
theory used here has been analyzed and motivated from a variety of
standpoints elsewhere \cite{Z2,NL}, we hope to make
clear its reasonableness on purely phenomenological grounds. 
We begin, then, from a zero-temperature theory of
d-wave paired spinons and charge $e$ bosons. The bosons 
can have a zero-temperature phase transition between condensed and
quantum disordered phases. We explore
here quantitatively the single-electron spectral function in the
$x=0$ spin-charge separated antiferromagnet ($AF^*$), the nodal liquid 
(to be identified with the pseudogap phase), and
the superconductor. 

\section{the model}

We briefly recapitulate the phase diagram of the cuprates in terms of
the $Z_2$ gauge theory introduced elsewhere \cite{Z2}. The theory
contains 
spinon and chargon degrees of freedom, coupled to a $Z_2$ gauge
field in two spatial dimensions. We begin with the square lattice
Hamiltonian: 
\begin{eqnarray}
H &=& \sum_{<ij>} \hat{\sigma}^z_{ij} [-t_s
\hat{f}_{i\alpha}^{\dag}\hat{f}_{j\alpha} 
+ \Delta_{ij}\hat{f}_{i\uparrow}\hat{f}_{j\downarrow} - t_c
\hat{b}_i^{\dag}\hat{b}_j + H.c.] 
\nonumber \\
& & + U \sum_i [\hat{n}_i-(1-x)]^2 +\sum_i g\vec{N}\cdot \hat{\vec{S}^{\pi}_i}
\nonumber \\
& & - h\sum_{<ij>} \hat{\sigma}^x_{ij} - K \sum_{\Box} \prod_{\Box}
\hat{\sigma}^z_{ij}, 
\label{Z_2}
\end{eqnarray}
where the electron operator is a product of spinon and
chargon operators: $c_{i\alpha} = b_i f_{i\alpha}$. 
The term with coupling $K$ is allowed by symmetry and can 
arise from integrating out the very high energy
chargons, making this an effective theory of the low energy charge
degrees of freedom. The spinon pairing $\Delta_{ij}$ is taken to be
d-wave:
\begin{equation}
\Delta_{ij} = \left\{ \begin{array}{ll} +\Delta &
\mbox{along} \;\; \hat{x}, \\ -\Delta & \mbox{along} \;\; \hat{y}, \\
\end{array} \right. 
\end{equation}
and the spin operator is $\hat{\vec{S}^{\pi}} = \sum_{\goodsub{k}}
\hat{f}^{\dag}_{\goodsub{k}+\goodsub{\pi}} \vec{\sigma}
\hat{f}_{\goodsub{k}}$.    
$\vec{N}$ is the mean-field N\'{e}el order parameter and is
 non-zero only within the antiferromagnetic phase.
The $U$ term is a Hubbard-like interaction for $(1-x)$ chargons per
unit cell. 
At zero temperature and as a function of
$K/h$, the gauge field has a transition between confining and
deconfining phases \cite{Z2}. Deep within
the deconfining phase, we may set $\sigma^z_{ij} =1$ on all links and we
are left with decoupled spinons and chargons:
\begin{eqnarray}
H &=& \sum_{<ij>} [-t_s \hat{f}_{i\alpha}^{\dag}\hat{f}_{j\alpha}
+ \Delta_{ij}\hat{f}_{i\uparrow}\hat{f}_{j\downarrow} - t_c
\hat{b}_i^{\dag}\hat{b}_j + H.c.]  \nonumber \\
& & + U \sum_i [\hat{n}_i -( 1-x)]^2 + \sum_i \vec{N} \cdot
\hat{\vec{S}^{\pi}_i} . 
\label{separated} 
\end{eqnarray} 

Fluctuations of $\sigma^z$ can be taken into account
by considering vortices in the Ising gauge field which have been
dubbed ``visons''. (A plaquette which contains a vison has
$\prod_{\Box} \sigma_{ij}^z = -1$.) The deconfining phase of the $Z_2$
gauge 
field is characterized by a gap to these vison excitations and, as we
see above, the electron degrees of freedom are fractionalized in this
phase. The
zero temperature confining phase of the $Z_2$ gauge field is a
condensate of these vison excitations and ``glues together''
spinons 
and chargons to form electrons. A ``quantum confinement critical
point'' separates these two zero-temperature phases, as discussed
elsewhere \cite{QC}. At finite temperatures above the fractionalized 
zero-temperature phase, we expect vison excitations to exist in two
dimensions, leading to interactions between the chargons and
spinons. However, at temperatures much smaller than the vison gap
$T^{vison}$ 
(which in the simplest theories of the quantum critical point would
be of the same order as the pseudogap temperature,  $T^*$ ), we
expect the low energy degrees of freedom to be those of the
fractionalized phase: spinons and chargons weakly interacting through
Doppler shift terms, which we ignore. 
  
We briefly discuss the phases shown pictorially in Figure
\ref{phases}.  
In the $x_c<x<x_{QCP}$ range, starting at temperatures much less than
the 
vison gap and lowering the temperature, the bosonic chargons should go
from a phase where they are 
phase incoherent to one where they are condensed. Below
the chargon condensation temperature, the system is superconducting;
this is $T_c$. Throughout, the spinons maintain a d-wave
pairing (presumably due to antiferromagnetic interactions of strength
$J$) and experience no phase transition, but rather a crossover at
their pairing scale, $T^*$.
Starting instead at zero-temperature and zero doping,
we are in a spin-charge separated phase which is also
an antiferromagnetic Mott insulator with long-range N\'{e}el order. Upon
increasing the doping, staying at zero-temperature, we presumably
enter a complicated charge-ordered insulating state of the chargons
and destroy the long-range N\'{e}el order of the spinons. This is the
zero-temperature 
phase believed to dominate the pseudogap region. We expect impurities
and thermal fluctuations to destroy static charge order,  but
inhomogeneous 
effects could still be an important high-energy presence, leading to
e.g, stripes. As the
doping is further increased, the chargons presumably condense at
zero-temperature into a superconducting state. 
After the destruction of N\'{e}el
order, the spinons are qualitatively the same in this doping
range and maintain a d-wave gap of order $T^*$. Throughout this
zero-temperature region, the chargons and spinons are decoupled (since
we are to the left of $x_{QCP}$). At $x=x_{QCP}$, the Ising gauge field
becomes confining and the chargons and spinons are bound together
to form electrons, presumably in a Fermi liquid phase.   

We turn our attention now to the spectral function defined in terms of
the electron Green function:
\begin{equation}
A(\good{k},\omega) = -\frac{1}{\pi} Im G(\good{k},\omega). 
\label{A} 
\end{equation}
Since at temperatures well below the
vison gap, we expect a description of the system in terms of
free chargons and spinons to capture the low-energy physics,
we use the Hamiltonian in Equation \ref{separated}, which is
a sum of spinon and chargon Hamiltonians: $H(c^{\dag},c) \simeq
H_{b}(b^{\dag},b) + H_{f}(f^{\dag},f) $.
Within this construction, it is possible
to write the electron Green function as a product of chargon and
spinon Green functions:
\begin{eqnarray}
\label{G} 
G(\good{r},\tau) &=& \langle T_{\tau}
c(\good{r},\tau)c^{\dag}({\bf 0},0) \rangle , \\ 
&=& \langle T_{\tau}b(\good{r},\tau) b^{\dag}({\bf
  0},0)\rangle \langle 
T_{\tau}f(\good{r},\tau) f^{\dag}({\bf 0},0) \rangle , \\
&=& G_b (\good{r},\tau) G_f (\good{r},\tau), 
\end{eqnarray}
with $\tau = it$, the imaginary time.
The problem of calculating the spectral lineshape in spin-charge
separated phases now becomes one of calculating the spinon and chargon
Green functions. We consider these two degrees of freedom in turn,
discussing values of various parameters in each phase.

\subsection{Spinons}

We briefly describe the phases
of the spinon model. Consider first the $\vec{N} = 0 $ phase which
describes spinons with a d-wave paring amplitude,
$\Delta_{\goodsub{k}}$. In this 
spin-charge separated construction, superconductivity is dependent
only on the charge degrees of freedom. When the bosonic chargons are
condensed ($\langle b \rangle \neq 0$), we are in a BCS d-wave
superconductor and the spinons are simply neutralized BCS
quasiparticles. When the chargons lack phase coherence, we are in a 
phase with no superconductivity, but with a d-wave gap to any
excitation with spin $1/2$, called elsewhere the nodal liquid
\cite{NL}. When $\vec{N}
\neq 0$, spinon-antispinon pairs condense forming a state with
long-range antiferromagnetic order, but still containing free spinon
excitations above a gap (of order J)  
\emph{which are separated from the chargons} due to the vison
gap. This spin-charge separated antiferromagnet has been dubbed $AF^*$
\cite{QC}. 

The spinon piece of the Hamiltonian in Equation \ref{separated}
is quadratic in the spinon operators, and we may diagonalize it
using a Bogoliubov-type transformation. Setting
$g \vec{N} = N_0 \hat{z}$ and working in units of the lattice
constant, we obtain:
\begin{eqnarray}
H_f &=& \sum_{\goodsub{k}} E_{\goodsub{k}} \hat{a}^{\dag}_{\goodsub{k}
,\alpha}\hat{a}_{\goodsub{k} ,\alpha}, \\ 
\label{Espinon} 
E_{\goodsub{k}} &=& \sqrt{N_0^2 + \Delta^2_{\goodsub{k}} +
\epsilon^2_{\goodsub{k}}} ,\\ 
\epsilon_{\goodsub{k}} &=& -t_s(cos k_x  + cos k_y) ,\\
\Delta_{\goodsub{k}} &=& \Delta (cos k_x - cos k_y) , 
\end{eqnarray}
with,
\begin{eqnarray}
\hat{a}_{\goodsub{k} ,\alpha} &=& u_{\goodsub{k}} \hat{d}_{\goodsub{k} ,\alpha}
+ \alpha v_{\goodsub{k}}
\hat{d}^{\dag}_{-\goodsub{k} ,-\alpha},  \\
u^2_{\goodsub{k}} = \hlf\ + \hlf\ \cos\theta_{\goodsub{k}} &;&
    v^2_{\goodsub{k}} =\hlf\ - \hlf\ \cos\theta_{\goodsub{k}} , \\
\cos\theta_{\goodsub{k}} &=&
    \frac{\epsilon_{\goodsub{k}}}{\sqrt{\epsilon^2_{\goodsub{k}} +
    \Delta^2_{\goodsub{k}}}}, 
\end{eqnarray}
where 
\begin{eqnarray}
\hat{d}_{\goodsub{k} ,\alpha} &=& A_{\goodsub{k}} \hat{f}_{\goodsub{k} ,\alpha}
+ \alpha B_{\goodsub{k}} \hat{f}_{\goodsub{k} + \goodsub{\pi} ,\alpha},\\  
A^2_{\goodsub{k}} = \hlf\ + \hlf\ \cos\phi_{\goodsub{k}} &;&
B^2_{\goodsub{k}} = \hlf\ - \hlf\ \cos\phi_{\goodsub{k}}, \\
\cos\phi_{\goodsub{k}} &=& \frac{\sqrt{\epsilon^2_{\goodsub{k}}+
\Delta^2_{\goodsub{k}}}}{E_{\goodsub{k}}},  
\end{eqnarray}
is a Hartree-Fock-type spin density wave operator at momentum
{\boldmath $\pi$}  
appropriate to commensurate antiferromagnetic order
with sublattice magnetization $N_0$ \cite{gros}. Note that
when $N_0 = 0$, the Hamiltonian for $\hat{a}_{\goodsub{k}}$ is the
same as the effective BCS Hamiltonian for a d-wave
superconductor. Indeed, when the chargons condense, these become the
Bogoliubov d-wave quasiparticles. 

At zero temperature, we have for the spinon correlation function:
\begin{equation}
\langle f^{\dag}_{\goodsub{k} \alpha} f_{\goodsub{k} \beta} \rangle =
\frac{\delta_{\alpha,\beta}}{2} \left( 1-
\frac{\epsilon_{\goodsub{k}}}{E_{\goodsub{k}}}  \right) . 
\end{equation}
We see that the spinon spectrum now has a gap of $N_0$ at
$k_x=k_y=\frac{\pi}{2} $, as we would
expect in the N\'{e}el state, as well as a d-wave gap whose maximum
is $2\Delta$ . 
We expect this spinon theory to work qualitatively at all
temperatures well below $T^*<T^{vison}$, where the spinons are
strongly paired and the low energy degrees of freedom are
fractionalized. In 
particular, one should note that 
in the absence of chargon-spinon interactions, the spinons do not
notice $T_c$.

The parameters $t_s$ and $\Delta$ can be set by the experimentally
determined ratio:
\begin{equation}
\frac{t_s}{\Delta} = \frac{v_f}{v_{\Delta}} ,
\end{equation}
which ranges from $\sim 14$ in YBCO to $\sim 20$ in BSCCO near optimal
dopings \cite{TF}. We expect this ratio to be of this order throughout the
pseudogap phase.
At zero doping, $|\vec{N}|$ is on the order of $J$.

\subsection{Chargons}

Once liberated from their fermionic statistics, the charge degrees of
freedom behave as bosons of charge $e$ hopping on a 2d square
lattice, as in Equation \ref{separated}. At half-filling and zero
temperature, we expect that in the limit $U/t_c >> 1$, the system
forms a Mott insulator, while in the limit $t_c/U >> 1$, the bosons
form a superfluid. This can be
described by the 2+1-dimensional quantum XY model which has two
phases, a superconducting 
phase and a quantum disordered, Mott insulating phase.
Being concerned primarily with ``normal state''
(i.e. non-superconducting) properties, consider the insulating phase
where $U/t_c >>1$. Excitations of this phase are doubly-occupied sites
which are ``massive'' (i.e. gapped) and may propagate. 
These excitations as well as the excitations within the superfluid
phase 
are well-described by the soft-spin continuum Landau-Ginzburg
action, replacing the chargon operator with a complex field. 
\begin{eqnarray}
\hat{b}_i &\rightarrow&  b(\good{r}), \\
{\cal L}_b &=& \hlf\ |\partial_{\tau} b |^2 + \frac{v^2}{2}
|\nabla b|^2 + \frac{\mu}{2} |b|^2 + u (|b|^2)^2. 
\label{softspin} 
\end{eqnarray}
When $\mu > 0$, the bosons are quantum disordered and the chargon
system is insulating. When $\mu < 0$ , the chargons
condense, forming a superconductor with $|\langle b \rangle|^2 =
\mu/4u = n_0 $, where $n_0$ is the condensate density. Fluctuations
around this new 
minimum are described (to quadratic order) by the action:  
\begin{eqnarray}
b(\good{r},t) &=& \langle b \rangle + \tilde{b}(\good{r},t), \\
\tilde{b} &=& \tilde{b}_1 + i \tilde{b}_2, \\
{\cal L}_b &=&  \hlf\ (\partial_{\tau} \tilde{b}_1)^2 + \frac{v^2}{2}
(\nabla \tilde{b}_1)^2 + \frac{M^2}{2} (\tilde{b}_1)^2 \nonumber \\ 
& & + \hlf\ (\partial_{\tau} \tilde{b}_2)^2 + \frac{v^2}{2}(\nabla
\tilde{b}_2)^2, \\ 
\label{superconductor}
\mbox{with}, & & M^2 = -2\mu .
\end{eqnarray}
Starting in the superconducting phase and increasing the chemical
potential toward $\mu =0$, the order parameter (and therefore the
condensate density) vanishes at the transition. 

Away from half-filling in the presence of long-range Coulomb
interactions or disorder, we expect the unoccupied sites to form some
crystal. Even in the case of zero disorder, the underlying lattice
makes characterization of this phase difficult. 
One way to gain intuition for this regime is by reformulating the
problem in terms of vortices in the chargon phase \cite{vortices}. 
On physical grounds, we expect that the strong coupling of the
charge degrees of  freedom will lead to complicated charge ordered
states at zero 
temperature when the number of bosons is incommensurate with the
underlying lattice, and that with increasing doping,
the system should eventually pass into a zero-temperature
superconducting state. 
The location of the transition, $x_c$, and the nature of the exact
ground state for $x < x_c$ will depend sensitively on the chargon
interactions and lattice commensurability effects.  
Lacking a more detailed theory of chargon solidification away from
half filling in a 
fractionalized phase, we will use the XY model defined above to
describe the low-energy degrees of freedom at low temperatures in the
fractionalized phases. Our main motivation is simplicity: the 2+1d XY
model contains both a quantum disordered and a superconducting phase
of bosons, as the more correct theory of the $x>0$ boson system
should. Although this description is obviously inadequate to
describe the zero-temperature phases away from half-filling as well as
the detailed critical properties of the transition, we note that
for a perfectly clean 
system, the physics at length scales shorter than the mean hole
spacing should be those of the half-filled system. At dopings of, say, 5
percent, this length is about 5 lattice spacings. In the corresponding
energy range, the 2+1 XY model should capture the correct physics. We
note that ARPES is an intermediate energy probe, although the energies
corresponding to moderate dopings may still be too high. Since we are 
concerned here with general features of spectral function in each
phase, we work with this phenomenological
Landau-Ginzburg model, hoping to capture the correct physics.

Therefore, in the charge disordered (nodal liquid) phase at temperatures
much smaller than 
the vison gap, we use Equation \ref{softspin} and find for the chargon
correlation function (setting $\mu =m^2 >0 $ and ignoring $u$ to
lowest order):
\begin{eqnarray}
\label{charge1}
\langle b^{\dag}_{\goodsub{k}}  b_{\goodsub{k}} \rangle &=&
\frac{1}{\omega_{\goodsub{k}}} , \\ 
\omega^2_{\goodsub{k}} &=& m^2 + v^2 |\good{k}|^2.
\label{charge2} 
\end{eqnarray}

We briefly discuss the parameters in the model, $m$ and $v$.
In the underdoped regime near
the critical point, $x_c$, we expect the chargon gap to be quite
small, while at half-filling (in the parent 
insulator) the chargon gap is rather large, the charge
gap for these materials being on the order of electron volts.
Very little can be said about the velocity $v$ without a more detailed
microscopic theory. Working in units where \good{k} is a dimensionless
wavenumber, $v$ is an energy scale, and we take it to be
moderately larger than the spinon kinetic scale, $t_s$, 
(but of the same order) effectively giving the charge
excitations a larger bandwidth than the spinons.

In the ordered phase ($\mu < 0$ ), the bosons are superconducting and
at $T=0$ the chargon-chargon correlation function has the property:
\begin{equation}
\begin{array}{c} \mbox{lim} \\ \goodsub{r}'-\goodsub{r} \rightarrow
\infty \end{array} 
\langle b^{\dag}(\good{r},t) b(\good{r}',t') \rangle = |\langle b
\rangle|^2  = n_0,
\end{equation}
the condensate density. The quantity $\langle b \rangle $ is precisely
the order parameter of the superconductor. 
At dopings $x>x_c$, as temperature increases within the superfluid
phase, we expect phase 
fluctuations to reduce this quantity, eventually causing it to vanish
at $T=T_c(x)$. At zero temperature within the superfluid phase, as
doping is decreased it will vanish at $x = x_c$. 
As discussed above, the details of the $T=0$ transition
will be governed by the universality class of the true
doping-dependent chargon theory. 

Within the superconducting phase, we may model the bosons with
Equation \ref{superconductor}, which results in the following general
form for the chargon spectral function: 
\begin{equation}
\label{SCchargons}
\langle b^{\dag}(\good{r},t)b(\good{r}',t') \rangle = n_0(T) +\langle
\tilde{b}^{\dag}(\good{r},t)\tilde{b}(\good{r}',t') \rangle .
\end{equation}
The fluctuations of the chargon field will be dominated by the
detailed interactions between the chargons. 
In contrast to the Cooper pairs in a standard BCS superconductor, the
bosonic chargons should be strongly interacting, given that their
uncondensed phase is controlled by Mott-insulating physics.
However, we expect that at
energies larger than the condensation temperature, $T_c$,  the chargon
fluctuations should be the same as in the pseudogap state.  

\section{Spectral Function}

\begin{figure}
\epsfxsize=3.0in
\centerline{\epsffile{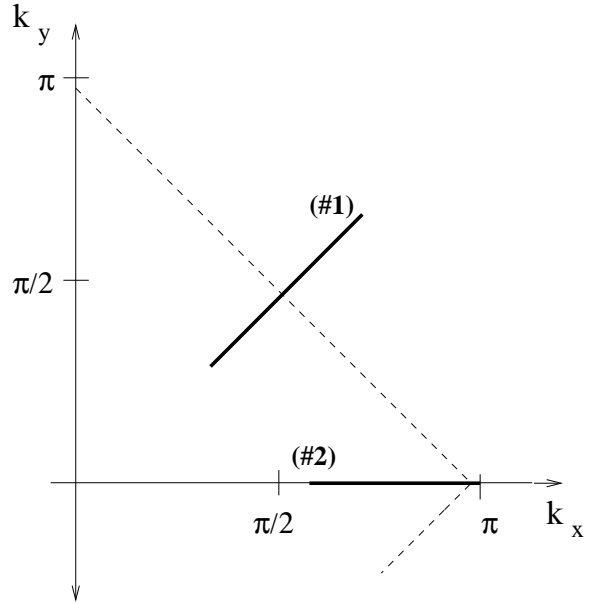}}
\vspace{0.15in}
\caption{Momentum cuts used for plots of $A_{-}(\good{k},\omega)$, showing
the approximate location of the ``Fermi surface'' for our model. Cut
\#1 (used for MDCs and EDCs)
is along the line $k_x = k_y$ near the nodal point, cut \#2 (used for
EDCs) is 
along $k_y =0$ near the antinodal point.}  
\vspace{0.15in}
\label{cuts}
\end{figure}

Given the spinon and chargon correlation functions, we can compute the
electron spectral function, assuming no interactions between chargons and
spinons, using the relations in Eqns. \ref{A} and \ref{G}. The result at
zero temperature is:
\begin{eqnarray} 
\label{spectral} 
 &A(\good{k},\omega)& = \nonumber \\
 &=&  \int_{\goodsub{q}}\left[ \langle f_{\goodsub{q}}
f^{\dag}_{\goodsub{q}} \rangle \langle b_{\goodsub{k}-\goodsub{q}}
b^{\dag}_{\goodsub{k}-\goodsub{q}} \rangle \delta(\omega -
\omega_{\goodsub{k}-\goodsub{q}} + E_{\goodsub{q}}) \right. \nonumber \\
  & &+ \left. \langle f^{\dag}_{\goodsub{q}}
f_{\goodsub{q}} \rangle \langle b^{\dag}_{\goodsub{k}-\goodsub{q}}
b_{\goodsub{k}-\goodsub{q}} \rangle \delta(\omega +
\omega_{\goodsub{k}-\goodsub{q}} + E_{\goodsub{q}}) \right] , \\
&=& A_{+}(\good{k},\omega) + A_{-}(\good{k},\omega).
\end{eqnarray} 
Because it measures electrons ejected from the sample, the ARPES
intensity (up to matrix element effects) measures the occupied part of
the spectral function, $A_{-}(\good{k}, \omega)$ \cite{MR}. 
At temperatures far below the vison gap, the assumption of no
chargon-spinon interactions should be valid. At
energies, $\omega$, larger than the temperature, the use of
zero-temperature results should be valid. Most of the ARPES data of
interest are done at temperatures below 100 K, which translates to
an energy of 10 meV or less, close to the resolution of the
instruments and certainly smaller than any features in the ``normal
state'' spectra. This justifies use of the zero-temperature
spectral function in our low-energy model. We therefore compare
this $A_{-}(\good{k},\omega) $ with the ARPES data 
in each of the following phases: $AF^*$, nodal liquid
(pseudogap), and d-wave superconductor. Although Equation
\ref{spectral} is quite simple, it 
nevertheless is not analytically integrable for arbitrary \good{k}
and $\omega$. In the following
sections, we present the results of numerical integrations of this
function and plot the resultant $A_{-}(\good{k},\omega) $ at fixed
\good{k} (EDC) and at fixed $\omega$ (MDC), along the momentum cuts
shown in Figure \ref{cuts}. For the numerical integration, we
approximate the delta function in Eqn. \ref{spectral} by a Lorentzian
of small 
width (0.0125 eV) for the energy distribution curves. This leads to
small ``tails'' in these curves at small binding energies (near
turn-on).  Since we would like to explore the momentum distribution
curves at these small turn-on energies, we need to avoid measuring
mostly these Lorentzian tails. To this end, for the MDCs, we use
instead a box-like delta function:
\begin{equation}
\delta (x) = \left \{ \begin{array}{ll} 1/\epsilon & \mbox{for}
-\epsilon/2 < x < \epsilon ,\\ 0 & \mbox{else}, \\ \end{array} \right.
\end{equation} 
with $\epsilon = 10$ meV. Values  
of various parameters (such as $\Delta$ and $v$ ) will be given in
each section. 
Some ``Fermi surface'' properties of the spinons should be discussed.
While the pairing terms technically destroy any true
Fermi surface, the number of spinons at a given momentum still drops off
for $\epsilon_{\good{k}} > 0$ and is sensitive to the minimum of the
spinon dispersion, $E_{\goodsub{k}}$, which we can call $k_f$. 
Along the $(\pi,\pi)$ direction, this minimum occurs at
$\good{k} = (\pi/2,\pi/2)$, while along the $(\pi,0)$ direction, the
location of the minimum depends on the relative values of $t_s$ and
$\Delta$ and will be discussed in each phase. 

\subsection{(Fractionalized) Antiferromagnet}

In this phase, the spinons are particle-hole paired into an
antiferromagnet (with single spinons above the gap) and the
chargons are gapped into a Mott insulating phase. However, because
these two particles propagate as separate excitations, we expect an
electron injected into the system to ``decay'' or fractionalize into
these two constituents. We therefore expect the spectral function at
temperatures much lower than the vison gap 
 to be broad, without the delta-function peak at some \good{k} and 
$\omega $ which one finds when the underlying phase has electron-like
elementary excitations. 

\begin{figure}
\epsfxsize=3.5in
\centerline{\epsffile{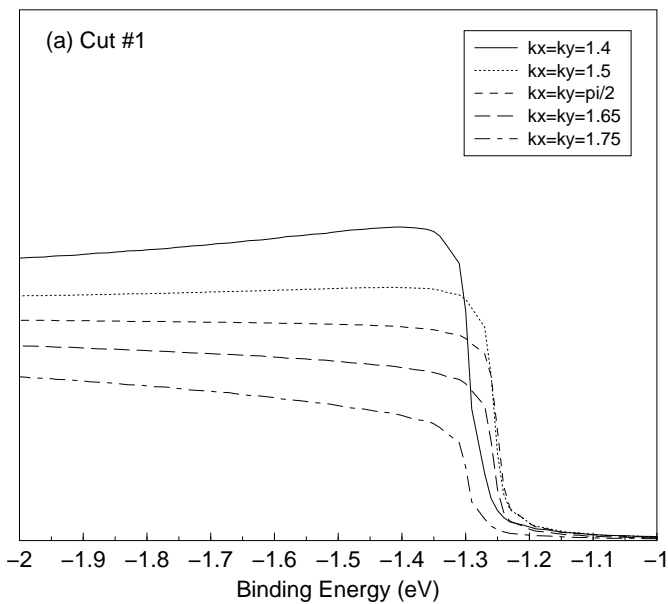}}
\vspace{0.10in}
\epsfxsize=3.5in
\centerline{\epsffile{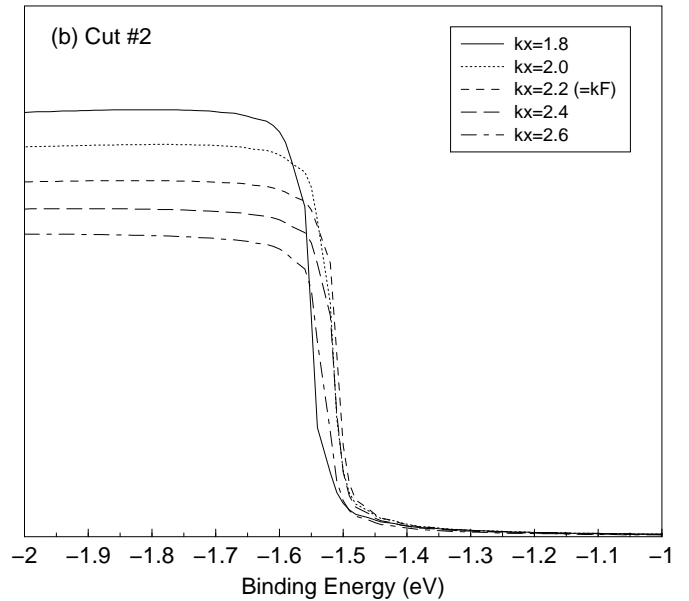}}
\vspace{0.10in}
\epsfxsize=3.5in
\centerline{\epsffile{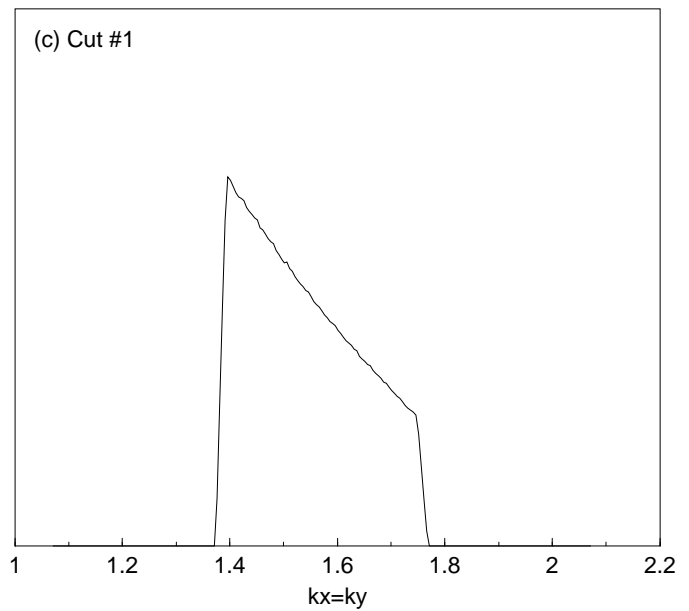}}
\vspace{0.10in}
\caption{$A_{-}(\good{k},\omega)$ at zero temperature in the $AF^*$ 
phase. Plotted are: EDCs along (a) cut \#1, and (b) cut \#2 and an MDC
(c) along cut \#1 at energy $ \omega = -1.30 $ eV.  The momentum space
cuts are shown in Figure \ref{cuts}.} 
\vspace{0.15in}
\label{AF}
\end{figure}

For the spinons, we
expect that both $\Delta$ and $N_0$ are of the order of $J \simeq
t_s/2$; we take
$t_s \simeq 0.5 eV$. The chargon gap, $m$,
is expected to 
be fairly large, on the order of an electron volt. With this 
in mind, we plot the electron spectral function in Figure \ref{AF} in
the $AF^*$ phase with  
$N_0= \Delta = 0.25 $ eV, $t_s= 0.5 $ eV, $v= 2.5 $ eV, and $m= 1 $
eV. The shapes of the 
curves are not sensitively dependent on any of these parameters.
For this ratio of $t_s$ to $\Delta$, the minimum of the spinon energy,
$E_{\goodsub{k}}$, along cut \#2 occurs at $k_x \simeq 2.2$. 
We plot the MDC along cut \#1, at the energy $\omega = -1.30$ eV. 
This is slightly larger than the minimum binding energy of $N + m =
1.25 $ eV.

A few features of these curves should be pointed out. First, all
EDCs are quite smeared, with no peaks. This indeed mimics one of
the features of ARPES data on the undoped compounds \cite{ZX}.
Detailed comparison with the binding energies in the ARPES data for
the undoped  compound is complicated by the fact that determining
the ``Fermi level'' of these compounds is not as straightforward as in
the doped materials. In the
case of our EDC plots, the binding energy should be used to note that the
leading 
edge along cut \#1 indeed has a smaller gap than
that along cut \#2. In fact, the difference between the two is just
the factor of $2\Delta$. That this feature of the leading edge will
track $\Delta_{\good{k}}$, the d-wave 
gap, can be seen from Equation \ref{spectral}. This is consistent with
the experimentally determined ``remnant Fermi surface'' with a d-wave
character found in the undoped compounds \cite{ZX}.
In contrast to the EDCs, the MDC shows a sharp feature. The detailed
shape of the MDC curves may be influenced by the specifics of the
model used.  

\subsection{Pseudogap (Nodal Liquid)} 

Because this zero temperature phase is also fractionalized, we expect
broad spectral functions in this region.
The spinons are paired into d-wave singlets, leading to the
spin-gap. To be precise, we calculate the spectral function at
zero-temperature using the XY model described earlier. 
We expect a low-energy theory of quantum
disordered chargons to work qualitatively for the entire pseudogap
region, provided $T << T^{vison}$. At finite temperatures, the
zero-temperature spectral function can only be expected to capture
features at energies larger than T. We would therefore like to compare
our spectral function in this phase with ARPES data in the underdoped
compounds at $T^*>>T>T_c$. The chargons in this region should be
dominated by their zero-temperature critical point. Here, we use
the critical 2+1d XY theory for the chargons described previously,
again noting that this will not describe in detail the true
finite-doping critical point 
but will hopefully give an adequate effective theory for the
low-energy excitations.

As an illustrative calculation, we may analytically perform the
convolution integral in Equation \ref{spectral} for $\good{k} =
(\pi/2,\pi/2)$ and 
$\good{k} = (\pi,0)$ at small $\omega $, exactly at the XY
critical point, $m=0$, of Eqns.\ref{charge1}-\ref{charge2}.
Equation \ref{spectral} reads:
\begin{eqnarray}
& & A_{-}(\good{k},\omega) = \nonumber \\
& & \int \frac{d^2q}{(2\pi)^2 4\omega_{\goodsub{q}}} \left(1-
\frac{\epsilon_{\goodsub{k-q}}}{E_{\goodsub{k-q}}}
\right)\delta (\omega+ E_{\goodsub{k-q}} + \omega_{\goodsub{q}}).
\end{eqnarray} 
At the node, the spinon spectrum may be linearized for small momentum
and we find (after rotating to momenta parallel and perpendicular to
the nodal direction and setting $t_s = \Delta = \overline{v}$ for
simplicity): 
\begin{eqnarray}
& & A_{-}[\good{k}=(\pi/2,\pi/2),\omega \,\, \mbox{small}] \simeq \nonumber \\
& & \simeq \int \frac{d^2 q}{(2\pi)^2} \frac{1}{4vq} \left( 1 -
\frac{\overline{v} q_x}{\overline{v} q} \right) \delta(\omega + 
\overline{v} q + vq) \\
& & \simeq  \frac{1}{8\pi v}\int_0^{\infty}dq \, \delta[\omega +
(v+\overline{v} )q] = \frac{1}{8\pi v(v+ \overline{v})} \theta(-\omega).
\end{eqnarray} 
At the antinode, the spinon spectral function is quadratic above the
gap, $E_{\goodsub{k-q}} \rightarrow \tilde{E}_q = 2\Delta \sqrt{1-\hlf
q^2}$,  and we find:
\begin{eqnarray} 
& & A_{-}[\good{k}=(\pi,0),\omega \simeq \Delta] \simeq \nonumber \\
& & \simeq \int \frac{d^2 q}{(2\pi)^2} \frac{1}{4vq} \left( 1 +
\frac{v_F[q_x^2 - q_y^2]}{2 \tilde{E}_q } \right) \delta(\omega + vq +
\tilde{E}_q ) \\
& & \simeq  \frac{1}{8\pi v}\int_0^{\infty}dq \, \delta[\omega +
vq + 2\Delta + O(q^2)] \\
& & \simeq \frac{1}{8\pi v^2} \theta(-\omega - 2\Delta).
\end{eqnarray} 
We see that at these two particular points in \good{k}-space, the
electron spectral function turns on like a step function, not a peak,
in qualitative agreement with the ARPES results.

\begin{figure}
\epsfxsize=3.5in
\centerline{\epsffile{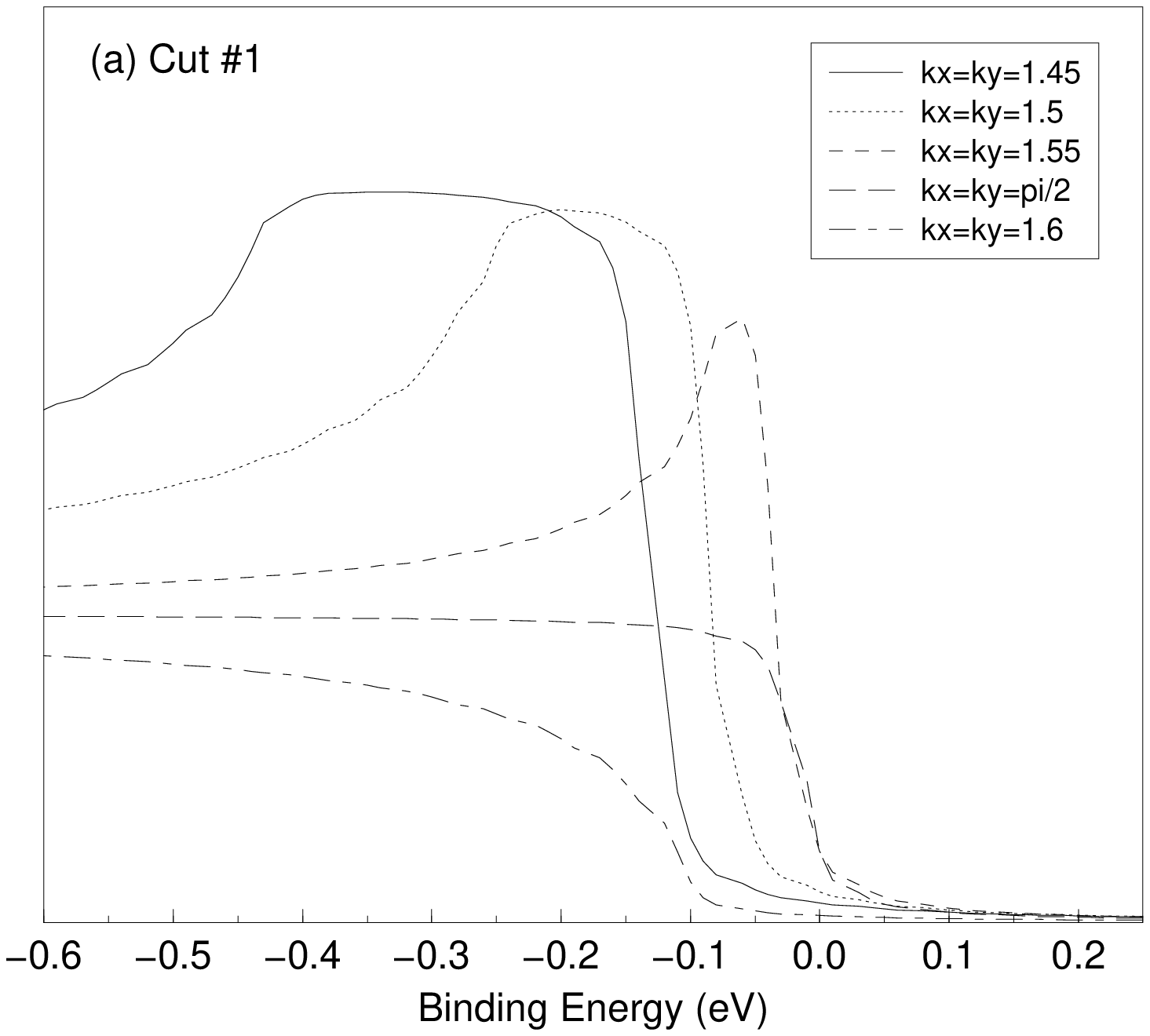}}
\vspace{0.10in}
\epsfxsize=3.5in
\centerline{\epsffile{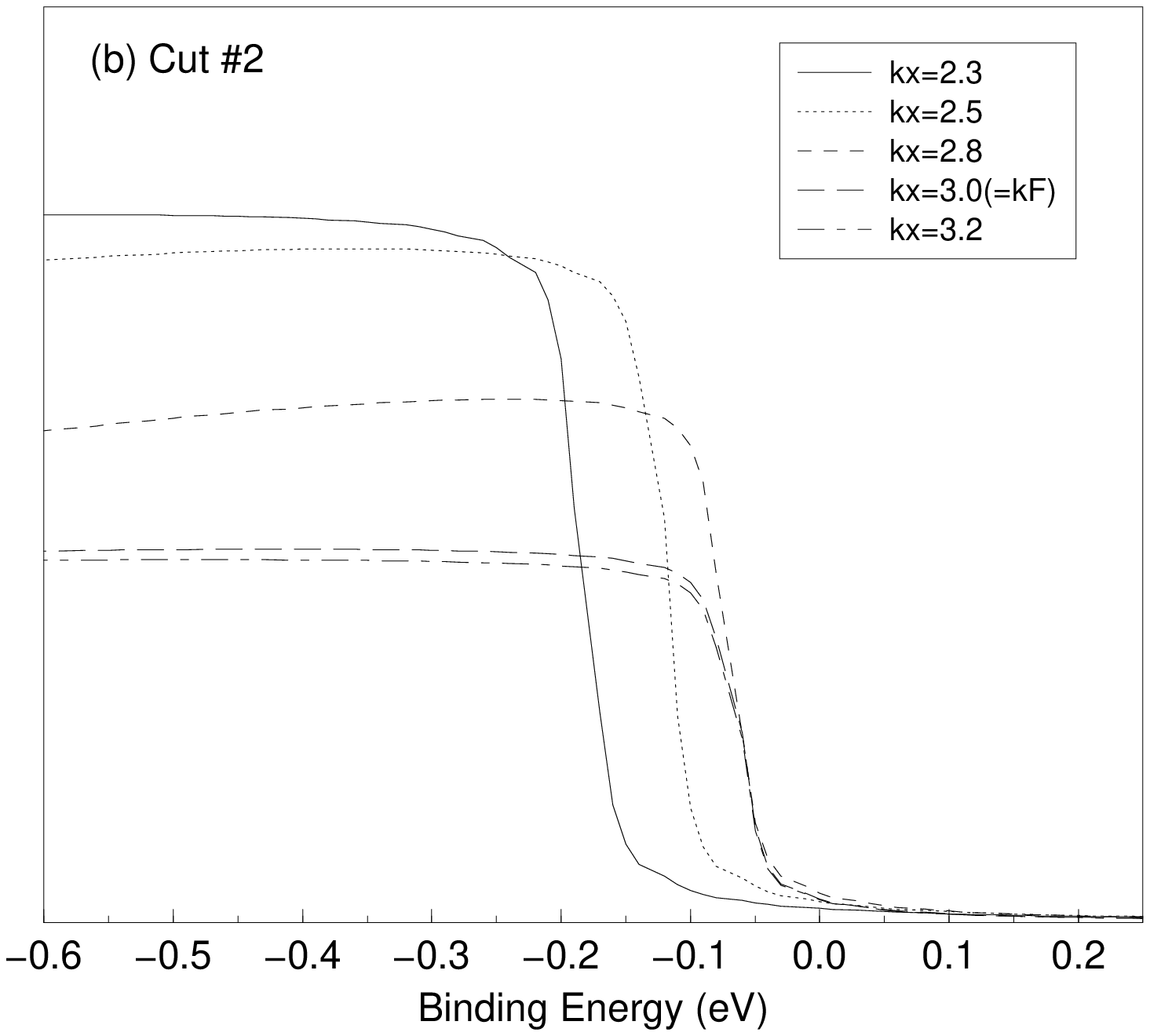}}
\vspace{0.10in}
\epsfxsize=3.5in
\centerline{\epsffile{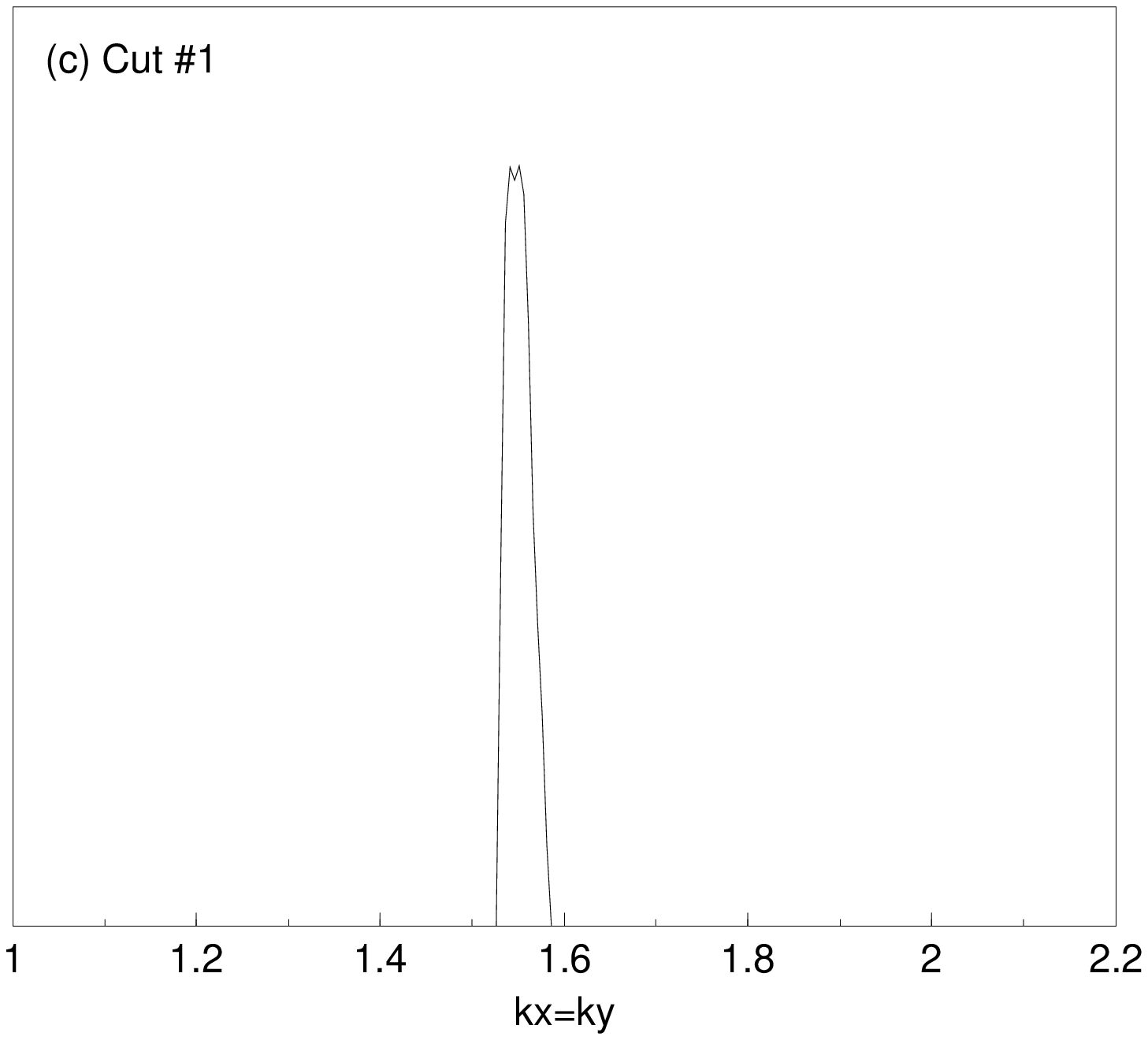}}
\vspace{0.10in}
\caption{$A_{-}(\good{k},\omega)$ in the pseudogap phase. Shown are:
EDCs along (a) cut \#1, and (b) cut \#2 and an MDC (c) along cut \#1
at an energy of $\omega = -40$ meV. The momentum space cuts are shown
in Figure \ref{cuts}.} 
\vspace{0.15in}
\label{NL}
\end{figure}

To obtain the electron spectral function at other \good{k} and
$\omega$, we resort to numerical integration.
In the underdoped region, values of $t_s/\Delta $ vary from
compound to compound, but are of order 10. 
With this in mind, we set $t_s = 0.5$ eV, $\Delta= 25$ meV, $v= 2.5$ eV.
Also, for the purposes of
performing the integration with the Lorentzian delta-function
approximation (see discussion at the beginning of section III), we
regularize the chargon spectrum by 
adding a small mass, $m= 12.5$ meV, for the energy distribution curves
shown in Figure \ref{NL}(a) and (b). For the momentum distribution
curve shown in Figure \ref{NL}(c), the chargon mass is equal to zero.  
Again, we find no 
sensitive dependence on the exact values of these parameters. For this
value of $t_s/\Delta$, the
minimum value of the spinon energy, $E_{\good{k}}$, along cut \#2
occurs at $k_x \simeq 3.0$.  For the MDC along cut \#1, we use a
binding energy large enough that the width of the approximate
delta-function does not influence the width of the curve, $\omega =
-40 $ meV.

We wish to note the following features of the graphs in Figure
\ref{NL}.
Foremost, the EDCs are indeed quite smeared, even near the ``Fermi
surface'' 
crossings of the spinons, but more so in the $(\pi,0)$ direction than
in the nodal 
direction, where something peak-ish (though still quite broad) emerges
near 
$(\pi/2,\pi/2)$. Also, as we have seen analytically, the leading edge
in the $(\pi,0)$ direction 
never gets to zero binding energy but instead shows a \emph{gap} of
$2\Delta = 50 meV$. It should also be pointed out that as one moves along
either cut, both
sets of EDCs show the leading edge moving toward its minimum binding
energy and then losing weight and/or receding above $k_f$.
Of particular interest is the contrast between the EDCs and MDC along
cut \#1 (the nodal direction), where the MDC shows a very sharp peak at
the node while the EDCs are broad and often step-like. The noise at
the top of the MDC is a consequence of using a ``box-like'' 
$\delta$-function for this integration.

\subsection{D-Wave Superconductor}

At low dopings (where $T^{vison}>>T_c$) when we cool below $T_c$, the
bosonic chargons develop phase coherence and $\langle b \rangle $ is
non-zero. The single electron correlation function in this region then
has two pieces, in accordance with Eqns. \ref{G} and \ref{SCchargons}:
\begin{eqnarray}
G(\good{r},\tau) &=& |\langle b \rangle|^2 \langle
f(\good{r},\tau)f^{\dag}({\bf 0},0)\rangle  \nonumber \\
&+& \langle \tilde{b}(\good{r},\tau) \tilde{b}^{\dag}({\bf 0},0) \rangle 
\langle f(\good{r},\tau)f^{\dag}({\bf 0},0)\rangle ,
\label{SCG} 
\end{eqnarray} 
giving an occupied portion of the spectral function:
\begin{eqnarray}
A_{-}(\good{k},\omega) &=& n_0(T) \langle f^{\dag}_{\goodsub{k}}
f_{\goodsub{k}} \rangle \delta (\omega + E_{\goodsub{k}}) \nonumber \\
&+& \int_{\goodsub{q}} \langle f^{\dag}_{\goodsub{q}}
f_{\goodsub{q}} \rangle \langle \tilde{b}^{\dag}_{\goodsub{k}-\goodsub{q}}
\tilde{b}_{\goodsub{k}-\goodsub{q}} \rangle \delta(\omega +
\tilde{\omega}_{\goodsub{k}-\goodsub{q}} + E_{\goodsub{q}}) .
\label{scA-}
\end{eqnarray}
Technically, this form is only valid at zero-temperature. However, we
can see from the electron Green function in
Eqn. \ref{SCG} (which is valid at all temperatures much less than
$T^{vison}$) that throughout the superconducting phase at low dopings, we
expect a spectral function made up of a peak and a background. 

The
peak is a product of the condensate density, $n_0(T)$, and the spinon
spectral function. For a given value of \good{k}, the peak is
located at the BCS quasiparticle energy, $E_{\goodsub{k}}$. Indeed,
for non-interacting bosons at zero temperature, $\langle
\tilde{b}^{\dag} \tilde{b} \rangle =0$, and we reproduce the BCS
quasiparticle peak. In contrast with the bosons of BCS theory, we
expect the chargons to be strongly interacting, leading to a
non-zero background even at zero temperature \cite{helium}.
We note that the width of this peak in our simple theory is entirely 
determined by the width of the spinon spectral function.
Throughout the superconducting state, we expect the spinons
to act like a two-dimensional Fermi liquid, leading to a weak
\cite{nodalfermi} 
temperature-dependent width. To the
extent that the peak and the background are distinguishable objects, the
weight under this quasi-particle peak should be proportional to
the condensate density, 
\begin{equation}
\int_{peak} A_{-}(\good{k},\omega) = n_0(T) \int_{peak}
\delta(\omega+E_{\goodsub{k}}) = n_0(T) ,
\end{equation} 
and should vanish into the background as 
$T \rightarrow T_c$ from below, without appreciable broadening. 

A comment should be made here regarding the difference 
between the condensate density, $n_0$, and the superfluid stiffness,
$\rho_s$. While for non-interacting bosons these quantities are the
same, for interacting bosons they
are different, even at zero temperature. Besides the effect of
chargon-chargon interactions on 
these quantities, there is the important effect of the Doppler shift
coupling between the superfluid and the quasiparticles in the
superconducting state. For a d-wave superconductor, the coupling
between quasiparticles and condensate leads to the
well-known T-linear depletion of the superfluid stiffness for small
T. The penetration depth, because it measures the superfluid stiffness,
manifests this dependence near $T=0$. The condensate density, on the
other hand, is not directly coupled to the quasiparticles, and
therefore need not approach $T=0$ in the same manner as the superfluid
stiffness.  
 
\begin{figure}
\epsfxsize=3.5in
\centerline{\epsffile{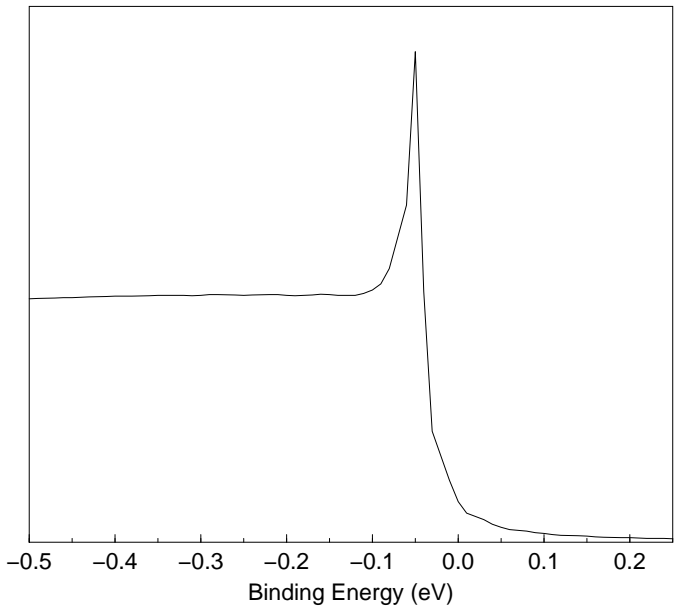}}
\vspace{0.15in}
\caption{For illustrative purposes only, this figure shows a
Lorentzian peak centered at $E_{\goodsub{k}}$ superimposed on the
nodal liquid $A_-(\good{k} ,\omega)$ for the ``Fermi surface''
crossing along cut \#2 from the previous section, $\good{k}=
(3.0,0)$.} 
\vspace{0.15in}
\label{SC_fig}
\end{figure}

The background in the spectral function comes from the second term in
Equation \ref{scA-}, and will be complicated by the exact nature of
chargon interactions.
At energies large compared to the condensation temperature ($\simeq$ 10
meV), we expect the spectral function to be that of the ``normal
state'' above $T_c$. 
At low energies, we have seen above that there
will be a sharp (resolution limited) peak in the
spectral function, with weight equal to the condensate density,
located at the spinon gap. It is only at intermediate energies,
(1-10meV), that the detailed physics of the chargons at their
charge-ordering critical point becomes important. 
In the superconductor for $T_c<<T^{vison}$, 
we therefore expect a sharp peak whose weight is given by the
condensate density, superimposed on a background which does not change
qualitatively as one moves from the superconductor to the pseudogap
phase above $T_c$. An illustration of the spectral function in the
superconducting phase is given in Figure \ref{SC_fig}. 

\section{Conclusions}

We have shown
here that the following aspects of
the ARPES data in the cuprate materials can be understood
by assuming spin-charge separation: 
1) the d-wave ``pseudogap'' seen above $T_c$, 
2) the lack of sharp quasiparticle peaks in the pseudogap phase,
3) the emergence of a very sharp quasiparticle peak below $T_c$,
4) the qualitative temperature and doping dependence of the weight
under this quasiparticle peak, as well as the existence within the
superconducting state of a background similar in shape to the pseudogap
spectra, and
5) the lack of sharp features in the undoped parent insulators as well
as the d-wave character of their ``remnant Fermi surface''.
We emphasize that
these results of ARPES in the undoped and underdoped compounds are
rather hard to account for within a conventional picture of
quasiparticles with the quantum numbers of electrons.

We are grateful to Leon Balents, J.-C. Campuzano, Steve Kivelson,
Doug Scalapino,
and Z.-X. Shen for helpful discussions. This research was generously 
supported by the NSF under Grants DMR-97-04005, DMR95-28578, and
PHY94-07194.

\end{multicols}


\begin{references}
\bibitem{vortices} The transition from superconductor to
fractionalized insulator at zero-temperature can be recast, through a
duality transformation, in terms of the proliferation and condensation
of $hc/e$ vortices. On a lattice with $1-x$  bosons per unit cell, these
vortices see $2\pi (1-x)$ flux through each plaquette. As the vortices
proliferate, this flux frustrates vortex motion, which we expect to
lead to spontaneous breaking of translational or rotational
symmetries.
\bibitem{helium} A well-studied case of this is superfluid $ ^4He$,
where even at zero temperature the condensate density, $n_0$, is much
less than the total density, $n$, due to interactions.
\end{references}
\end{document}